\documentclass[colorlinks]{goose-article}

\usepackage{lipsum}
\usepackage{ragged2e}

\newcommand{\detail}[1]{\textcolor{gray}{#1}}

\renewcommand{\detail}[1]{}

\renewcommand{\Re}{\text{Re}}
\renewcommand{\Im}{\text{Im}}

\title{Armouring of a frictional interface by mechanical noise}

\author{Elisa~El~Sergany}
\author{Matthieu~Wyart}
\author{Tom~W.J.~de~Geus}

\affil{
    Physics Institute,
    \'{E}cole Polytechnique F\'{e}d\'{e}rale de Lausanne (EPFL) Switzerland
}

\date{}

\hypersetup{pdfauthor={T.W.J. de Geus}}

\fancypagestyle{firstpage}
{
    \fancyhead[L]{Journal of Statistical Physics 191(10):134 (2024). \doi{10.1007/s10955-024-03339-z}}
    \fancyhead[R]{\thepage}
}

\begin{document}

\twocolumn[
    \begin{@twocolumnfalse}

        \maketitle

        \begin{abstract}
            \noindent
            A dry frictional interface loaded in shear often displays stick-slip.
            The amplitude of this cycle depends on the probability that a microscopic event nucleates a rupture and on the rate at which microscopic events are triggered.
            The latter is determined by the distribution of soft spots, $P(x)$, which is the density of microscopic regions that yield if the shear load is increased by some amount $x$.
            In minimal models of a frictional interface -- that include disorder, inertia and long-range elasticity -- we discovered an `armouring' mechanism by which the interface is greatly stabilised after a large slip event: $P(x)$ then vanishes at small argument as $P(x)\sim x^\theta$ \cite{deGeus2019}.
            The exponent $\theta$ is non-zero only in the presence of inertia (otherwise $\theta=0$).
            It was found to depend on the statistics of the disorder in the model, a phenomenon that was not explained.
            Here, we show that a single-particle toy model with inertia and disorder captures the existence of a non-trivial exponent $\theta>0$, which we can analytically relate to the statistics of the disorder.
        \end{abstract}

    \end{@twocolumnfalse}
]
\sloppy

\thispagestyle{firstpage}
\section{Introduction}

We study a class of systems in which disorder and elasticity compete, leading to intermittent, avalanche-type response under loading.
Examples include an elastic line being pulled over a disordered pinning potential and frictional interfaces~\cite{Fisher1998,Narayan1993,Kardar1998}.
When subject to an external load $f$, such systems are pinned by disorder when the load is below a critical value $f_c$.
At $f > f_c$, the system can move forward at a finite rate.
At $f = f_c$ the system displays a crackling-type response described by avalanches whose sizes and durations are distributed according to power laws.
This corresponds to a depinning transition.

A key aspect at the depinning transition is the distribution of soft spots \cite{Muller2015}.
If we define $x$ as the load increase needed to trigger an instability locally, then increasing the remotely applied load by $\Delta f$ will trigger $n_a \propto \int_0^{\Delta f} P(x) dx$ avalanches, with $P(x)$ the probability density of $x$.
The relevant behaviour of $P(x)$ therefore is that at small $x$.
Let us assume that $P(x) \sim x^\theta$ at small $x$, such that $n_a \propto (\Delta f)^{\theta + 1}$.

Classical models used to study the depinning transition consider overdamped dynamics.
In that case, it can be shown that $\theta = 0$ \cite{Fisher1998}.
This result is not true for certain related systems, including the plasticity of amorphous solids or mean-field spin glasses.
In these cases, due to the fact that elastic interactions are long-range and can vary in sign (which is not the case for the depinning transition where a local instability can only destabilise other regions: the interactions are strictly positive), one can prove that $\theta > 0$, as reviewed in~\cite{Muller2015,Rosso2022}.

\begin{figure}[htp]
    \subfloat{\label{fig:schematic:a}}
    \subfloat{\label{fig:schematic:b}}
    \centering
    \includegraphics[width=\linewidth]{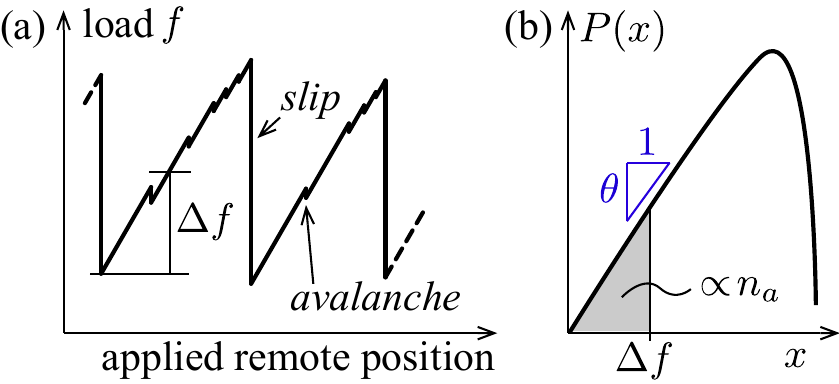}
    \caption{
        (a) Sketch of stick-slip response of an elastic interface with inertia to quasi-static loading in which the remote position in controlled.
        System-spanning ``slip'' events punctuate periods in which the interface is macroscopically stuck, but microscopic events (``avalanches'') do occur.
        The number of avalanches $n_a \propto (\Delta f)^{\theta + 1}$, with $\Delta f$ the load increase after a slip event.
        This can be linked to (b) the distribution of soft spots after a slip event.
        With $x$ the load increase needed to trigger an instability locally, we find that right after a large slip event, its distribution empirically scales like $P(x) \sim x^\theta$ at small $x$ as indicated (log-scale implied).
        We argue that $\theta > 0$ in the presence of mechanical noise, corresponding to an `armouring' of the interface such that it can build up potential energy after slip events, as sketched in (a).
    }
    \label{fig:schematic}
\end{figure}

Within the class of the depinning transition, we recently studied simple models of dry frictional interface \cite{deGeus2019,deGeus2022}.
We considered disorder and long-range elastic interactions along the interface that are strictly positive, as in the usual class of the depinning transition.
However, we included inertia, which turns out to have dramatic effects.
Inertia causes interactions to transiently overshoot and undershoot.
It thus generates a mechanical noise that lasts until damping ultimately takes place.
Remarkably, we found that right after system-spanning slip events, $\theta > 0$ \cite{deGeus2019} in the presence of inertia.
Intuitively, such an `armouring' mechanism results from the mechanical noise that destabilises spots close to an instability (i.e.~small $x$), thus depleting $P(x)$ at small argument.
This property is consequential: the number of avalanches of local instabilities triggered after a system-spanning slip event is very small.
As a consequence, the interface can increase its load when driven quasi-statically in a finite system, without much danger of triggering system-spanning slip events.
Such slip events eventually nucleate when disorder is unable to stop avalanches when the load is increased by a finite amount (whose typical value depends on $\theta$ through the number of avalanches, and on the probability that and avalanche can be stopped by disorder \cite{deGeus2019,deGeus2022}).
The interface therefore presents a stick-slip cycle, as sketched in \cref{fig:schematic}.
Thus, one of the central quantities governing the stick-slip amplitude is $\theta$~\cite{deGeus2019}.

Such quantity might be key in earthquakes.
Large earthquakes can be followed by rapid aftershocks, presumably due to slow creep and thermal effects not captured by our model. However, in the earthquake cycle, faults can then present a reduction in activity consistent with the armouring effect we propose \cite{Marone1998a}.
Indeed, earthquakes are acoustic waves which are clearly inertial objects. The presence of inertial is also clear in experiments on dry frictional interface, where acoustic emissions can be measured (e.g.~\cite{Passelegue2016a}).

Our previous model \cite{deGeus2019} divided the interface in blocks whose mechanical response was given by a potential energy landscape that, as a function of slip, comprised a sequence of parabolic wells with equal curvature.
We drew the widths $w$ of each well randomly from a Weibull distribution, such that its distribution $P_w(w) \sim w^k$ at small $w$.
We empirically found $\theta \simeq 2.5$ for $k = 1$ and $\theta \simeq 1.4$ for $k = 0.2$.

Here, we present a toy model that captures the qualitative relationship between microscopic disorder and $\theta$ after a system-spanning slip event.
This toy model assumes that regions of space (`blocks') do not interact during the last phase of a large slip event as the interface stops moving.
Although elastic interactions between these blocks are well-known to be key to the dynamics of the frictional interface, here we show that even a simple block in a potential is unlikely to stop in a shallow well, an effect described by some critical exponent.
In this most idealised view, we describe a region as a single particle that moves over a disordered potential energy landscape and that slows down due to dissipation.
We model this potential energy landscape by a sequence of parabolic potentials that have equal curvature $\kappa$ but different widths taken from $P_w(w)$, with $w$ the width of a parabola.
In this model, $x$ is thus proportional to the width of the well in which the particle stops, i.e.~$x = \kappa w / 2$.
Below we prove that for such a model, $P(x) \sim x^{k + 2}$ if $P_w(w) \sim w^k$ (at small argument in both cases).
This result explains both why $\theta>0$ and why this exponent in non-universal, as it depends on $k$ that characterises the disorder.
Although this prediction naturally does not match our previous observations quantitatively\footnote{
    As cited above: ``We empirically found $\theta \simeq 2.5$ for $k = 1$ and $\theta \simeq 1.4$ for $k = 0.2$''.
    Our toy model instead predicts $\theta \simeq 3$ for $k = 1$ and $\theta \simeq 2.2$ for $k = 0.2$.
} (as we neglect elastic interactions), qualitative agreement is already noticeable for such a simple model.
We support our argument with analytical proofs, and verify our conclusion numerically.
The generality of our argument suggests that the presence of a non-trivial exponent $\theta$ may hold in other depinning systems, as long as inertia is present.

\section{Model}

\begin{figure}[htp]
    \centering
    \includegraphics[width=\linewidth]{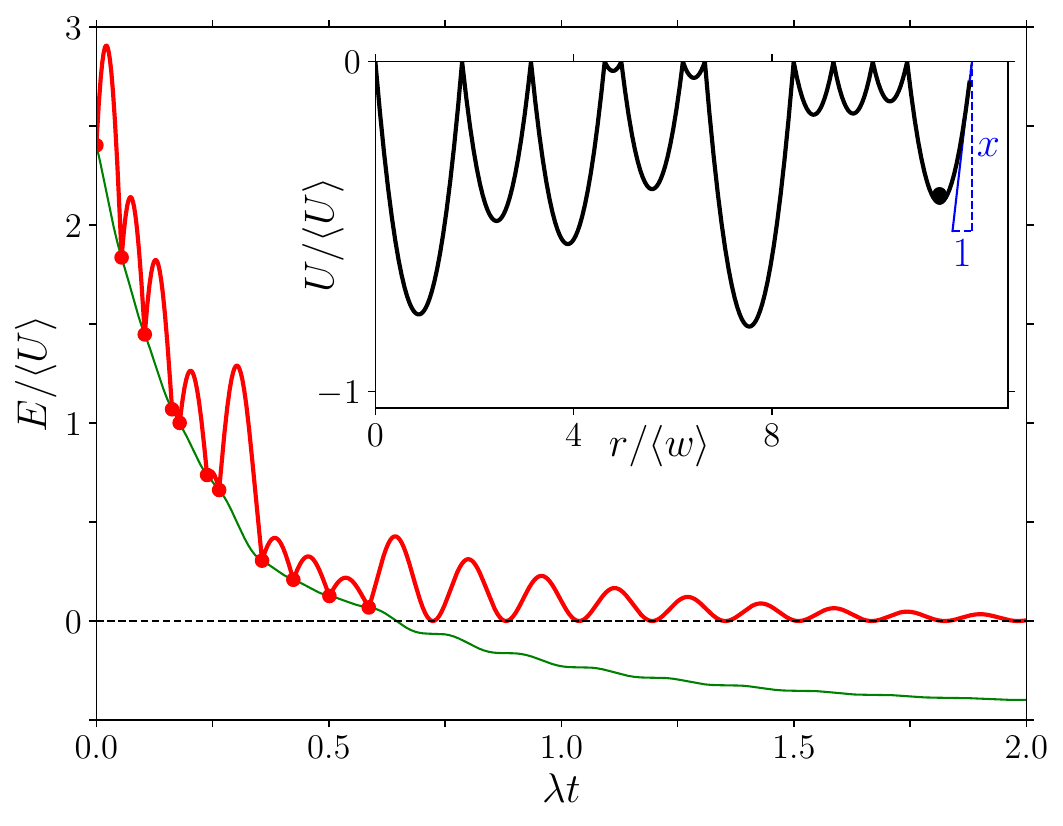}
    \caption{
        Evolution of the kinetic energy $E$ as a function of position $r$ (in red) of the ``free'' particle `thrown' into a potential energy landscape comprising a sequence of parabolic quadratic wells of random width $w_i$ (shown in the inset).
        Every entry into a new well is indicated using a marker (which corresponds to $\mathcal{E}$ in the text).
        A thin green line shows the evolution of the total energy (with the definition of the inset, it has the local minimum of the last well as arbitrary offset).
    }
    \label{fig:map}
\end{figure}

During a large slip event, all regions in space along an interface are moving but eventually slow down and stop.
For our toy model, we assume that regions in space do not interact during these final stages of a large slip event.
This allows us to model a single region in space as a particle of finite mass thrown into the potential energy landscape at a finite velocity.
In this simplest case, this particle is ``free'', such that it experiences no external driving and stops due to dissipation, see \cref{fig:map} (which contains a sketch of the model as inset).
Under these assumptions, a particular phase of the dynamics of a quasi-statically driven inertial elastic interface is approximated by the Prandtl-Tomlinson \cite{Prandtl1928,Tomlinson1929,Popov2014} model that describes the dynamics of one (driven) particle in a potential energy landscape.

The equation of motion of the ``free'' particle reads
\begin{equation}
    m \ddot{r} = f_e(r) -\eta \dot{r}.
    \label{eq:motion}
\end{equation}
with $r$ the particle's position, $m$ its mass, and $\eta$ a damping coefficient.
$f_e(r)$ is the restoring force due to the potential energy landscape.
We consider a potential energy landscape that consists of a sequence of finite-sized, symmetric, quadratic wells, such that the potential energy inside a well $i$ is given by $U(r) = (\kappa / 2) (r - r_{\min}^i)^2 + U_0^i$ for $r_\mathrm{y}^{i} < r \leq r_\mathrm{y}^{i + 1}$, with $w_i \equiv r_\mathrm{y}^{i + 1} - r_\mathrm{y}^{i}$ the width of the well, $\kappa$ the elastic constant, $r_{\min}^i \equiv (r_\mathrm{y}^{i} + r_\mathrm{y}^{i + 1}) / 2$ the position of the center of the well, and $U_0^i = \kappa w_i^2 / 8$ an unimportant offset.
The elastic force deriving from this potential energy is $f_e(r) \equiv - \partial_r U(r) = \kappa (r_{\min}^i - r)$.
With $\kappa$ constant, the landscape is parameterised by the distance between two subsequent cusps $w_i$, which we assume identically distributed (iid) according to a distribution $P_w (w)$.
We consider underdamped dynamics corresponding to $\eta^2 < 4 m \kappa$.
Within a well, the dynamics is simply that of a underdamped oscillator, as recalled in \cref{sec:proof:solution}.

\section{Stopping well}

\paragraph{Distribution}

We are interested in the width of the well in which the particle eventually stops.
Suppose that a particle enters a well of width $w$ with a kinetic energy $\mathcal{E}$.
The particle stops in that well if $\mathcal{E} < \mathcal{E}_c(w)$, with $\mathcal{E}_c$ the minimum kinetic energy with which the particle needs to enter a well of width $w$ to be able to exit.
The distribution of wells in which particles stop in that case is
\begin{equation}
    P_s(w) \sim P_w(w) P(\mathcal{E} < \mathcal{E}_c(w)) ,
    \label{eq:pstop}
\end{equation}
with $P_w(w)$ the probability density of well widths, and $P_s(w)$ the probability of well widths in which the particle stops.
Within one well, the particle is simply a damped harmonic oscillator, as has been studied abundantly.
In the limit of a weakly damped system, the amount of kinetic energy lost during one cycle is $\Delta E = \kappa w^2 (1 - \exp(-2 \pi / Q)) / 8$ with the quality factor $Q = \sqrt{4 m \kappa / \eta^2 - 1}$.
The minimal kinetic energy with which the particle needs to enter the well in order to be able to exist is thus $\mathcal{E}_c = \Delta E \propto w^2$ (see \cref{sec:proof:Ec} for a direct calculation of $\mathcal{E}_c$).
Furthermore, if $P(\mathcal{E})$ is a constant at small argument (as we will argue below), then
\begin{equation}
    P(\mathcal{E} < \mathcal{E}_c(w)) = \int_0^{\mathcal{E}_c} P(\mathcal{E}) \mathrm{d}\mathcal{E} \sim \mathcal{E}_c(w) \sim w^2 .
    \label{eq:pec}
\end{equation}
Therefore, the particle stops in a well whose width is distributed as
\begin{equation}
    P_s(w) \sim w^2 P_w(w) .
    \label{eq:rhos}
\end{equation}

\paragraph{Central result}

Once stopped, the force, $x$, by which we need to tilt the well in which the particle stopped in order for it to exit again is $x = \kappa w / 2$ \footnote{
    Without external forces, the particle ends in the local minimum -- the center of the well.
}, such that our central result is that
\begin{equation}
    P(x) \sim x^2 P_w(x).
\end{equation}
For example, if $P_w(w) \sim w^k$ at small $w$, we predict that
\begin{equation}
    \label{eq:px}
    P(x) \sim x^{2+k}.
\end{equation}

\paragraph{Energy at entry}

We will now argue that the density of kinetic energy with which the particle enters the final well is finite at small $\mathcal{E}$ (which is the only range of $P(\mathcal{E})$ that matters for the result above).
For one realisation, $\mathcal{E}$ results from passing many wells with random widths.
If its kinetic energy is much larger than the potential energy of the typical wells, it will not stop.
We thus consider that the particle energy has decreased up to some typical kinetic energy $E_0$ of the order of the typical potential energy $\kappa \langle w^2 \rangle / 8$.
If the particle exits the well, at exit it will have a kinetic energy $\mathcal{K} = E_0 - \Delta E(E_0, w)$.
For a given $E_0$ and distributed $w$ the kinetic energy with which the particle enters the next well is distributed as:
\begin{equation}
    P(\mathcal{E}) = \int dw \, P_w(w) \, \delta(\mathcal{K} (E_0, w) - \mathcal{E}).
\end{equation}
It thus implies that:
\begin{equation}
    P(\mathcal{E} = 0) = P_w(w^*) / \left| \partial_w \mathcal{K} \big|_{w = w^*} \right. ,
    \label{eq:rhoe0}
\end{equation}
where $w^*$ is the well width for which the particle reaches the end of the well with zero velocity, as follows from the equality$E_0 = \mathcal{E}_c(w^*)$.
By assumption, $P_w(w^*)>0$.
Furthermore, $\partial_w \mathcal{K} |_{w = w^*} = \kappa w^* /2 >0$ as we argue below and prove in \cref{sec:proof:Kw}.
Overall, it implies that $P(\mathcal{E} = 0)>0$, i.e.~$P(\mathcal{E})$ does not vanish as $\mathcal{E} \rightarrow 0$, from which our conclusions follow.

\begin{figure}[htp]
    \centering
    \includegraphics[width=\linewidth]{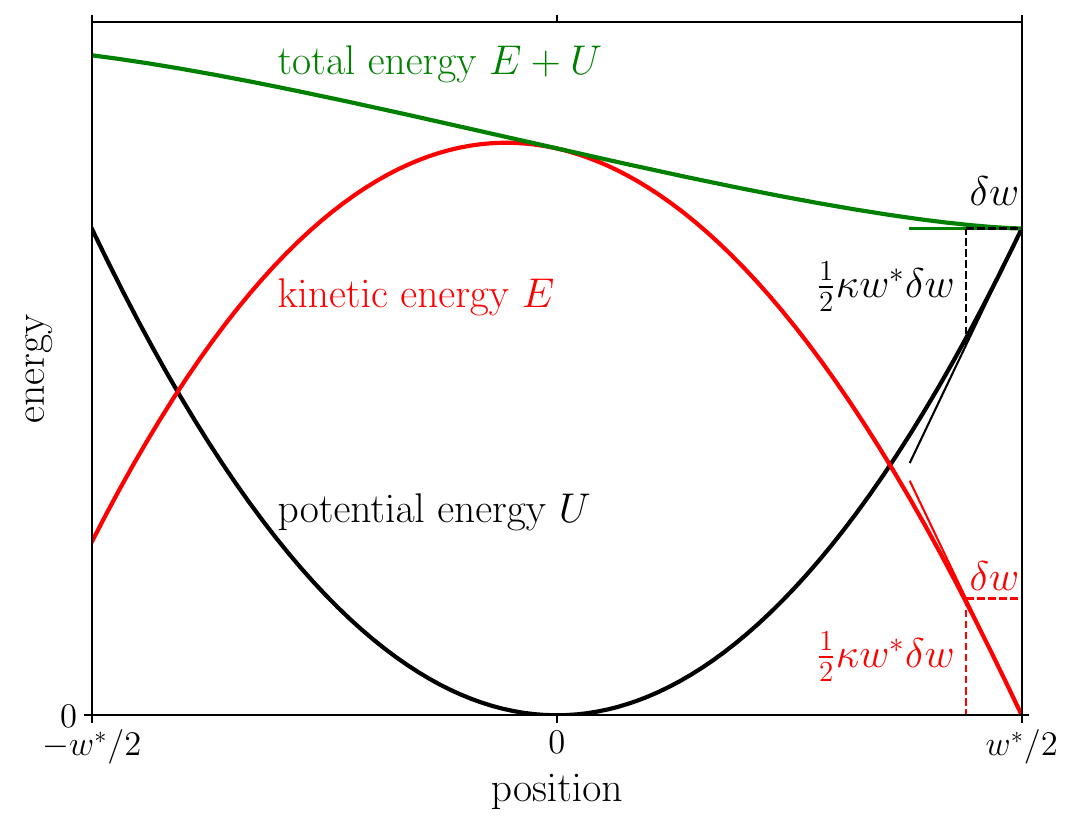}
    \caption{
        Evolution of the kinetic energy $E$ (red), potential energy $U$ (black), and total energy $E + U$ (green) for a particle that has entered a well of width $w^*$ with a kinetic energy $E_0 = \mathcal{E}_c(w^*)$ such that it stops just.
        Consequently, $\partial_r(E + U) |_{w^* / 2} = 0$, which can be decomposed in $\partial_r U |_{w^* / 2} = \kappa w^* / 2$ such that $\partial_r E |_{w^* / 2} = - \kappa w^* / 2$, as indicated using thin lines.
    }
    \label{fig:kappa}
\end{figure}

Here we give a simple argument for $\partial_w \mathcal{K} |_{w = w^*} = \kappa w^* / 2>0$.
Given $E_0 = \mathcal{E}_c(w^*)$, but an infinitesimally smaller well of width $w^* - \delta w$, the particle will enter the next well with a small kinetic energy $\delta \mathcal{K}$.
Because the velocity is negligible in the vicinity of $w^*$, the damping is negligible.
Therefore, $\delta \mathcal{K}$ is of the order of the difference in potential energy on a scale $\delta w$, $\delta U = U(w^*) - U(w^* - \delta w) \approx \kappa w^* \delta w / 2$, as we illustrate in \cref{fig:kappa}.
We thus find that $\partial_w \mathcal{K} |_{w = w^*} = \lim_{\delta w \rightarrow 0} \delta K / \delta w = \kappa w^* / 2$.

\section{Numerical support}

\paragraph{Objective}

We now numerically verify our prediction that $P(x) \sim x^{k+2}$ (\cref{eq:px}).
We simulate a large number of realisations of a potential energy landscape constructed from randomly drawn widths (considering different distributions $P_w(w)$) and constant curvature.
We study the distribution of stopping wells for a ``free'' particle that is `thrown' into the landscape at a high initial velocity such that particle transverses many wells before stopping (i.e.\ the kinetic energy with which the particle is thrown is much larger than $\mathcal{E}_c(\langle w \rangle)$).

\paragraph{Map}

We find an analytical solution for \cref{eq:motion} in the form of a map.
In particular, we derive the evolution of the position in a well centered at $r = 0$, based on an initial position $r_0 = -w / 2$ and velocity $v_0$ in \cref{sec:proof:solution}.
This maps the velocity $v_0$ to an exit velocity at position $r = w / 2$.
That exit velocity then corresponds to the entry velocity of the next well, etc.

\paragraph{Stopping well}

\begin{figure}[htp]
    \centering
    \includegraphics[width=0.5\textwidth]{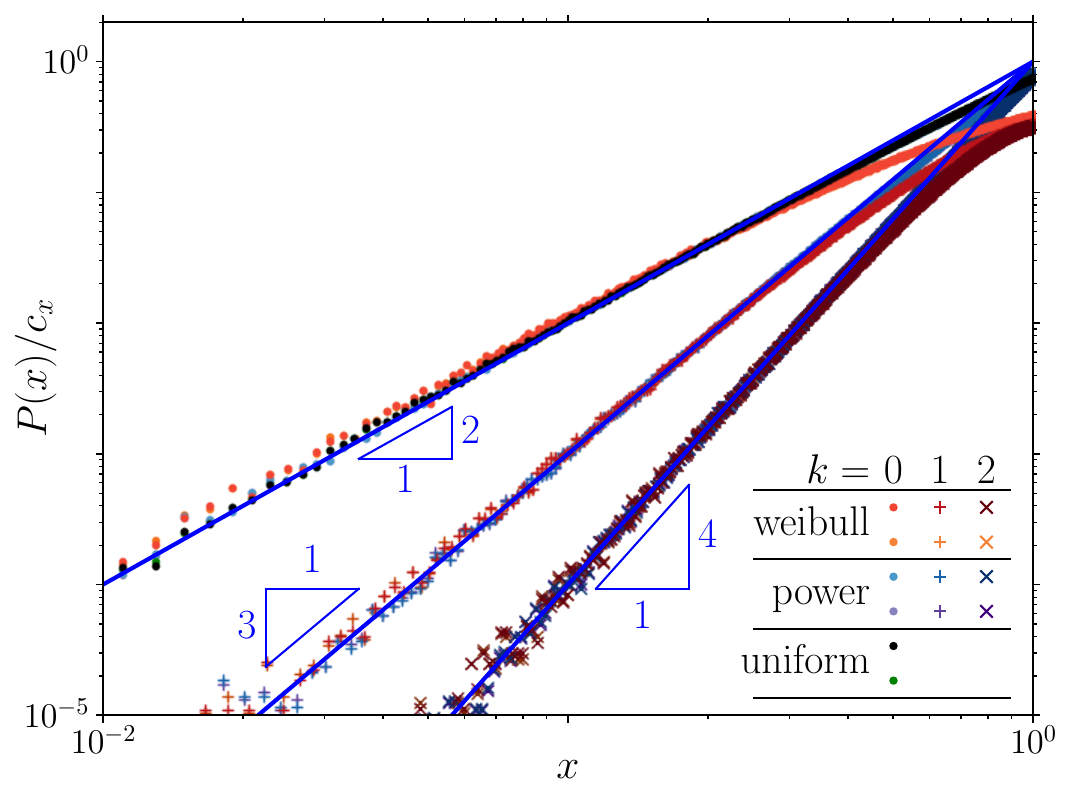}
    \caption{
        Width of the stopping well, $x$, for different $P_w(w)$: a uniform, Weibull, and powerlaw distribution, that scale as $P_w(w) \sim w^k$ at small $w$, as indicated in the legend (the bottom row for each distribution corresponds to perturbing the dynamics with random force kicks, tilting individual wells by a force $F = \mathcal{N}(0, 0.1)$, with $\mathcal{N}$ the normal distribution; the top row corresponds to $F = 0$).
        To emphasise the scaling, the distributions have been rescaled by a fit of the prefactors: $P(x) = c_x x^{k + 2}$.
        Furthermore, we use $m = \kappa = 1$, $\eta = 0.1$, $v_0 = \mathcal{N}(100, 10)$, and $\langle w \rangle \approx 1$.
    }
    \label{fig:pdf_x}
\end{figure}

\begin{figure}[htp]
    \centering
    \includegraphics[width=\linewidth]{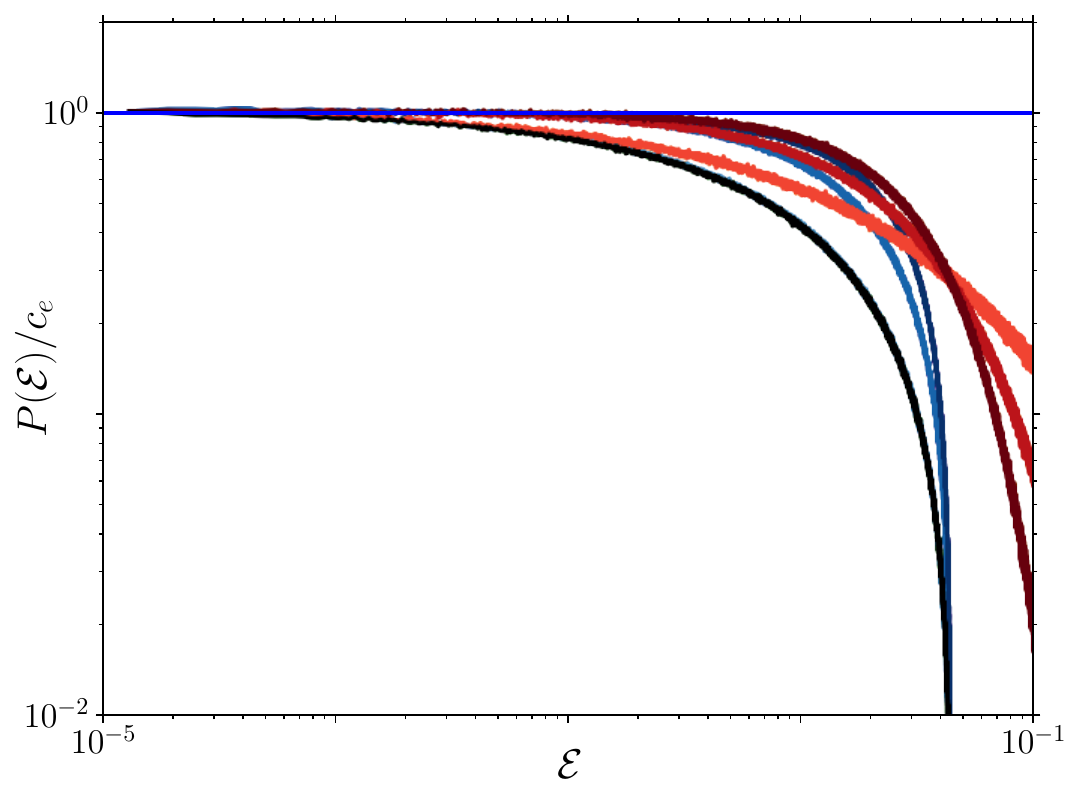}
    \caption{
        The kinetic energy with which the particle enters the well in which it stops for different realisations, $P(\mathcal{E})$, normalised by its prefactor $c_e$ (that is here simply the density of the first bin).
        See \cref{fig:pdf_x} for legend.
    }
    \label{fig:E}
\end{figure}

We record the width of the stopping well, $x$, and the kinetic energy $\mathcal{E}$ with which the particle enters the final well.
We find clear evidence for the scaling $P(x) \sim x^{k + 2}$ in \cref{fig:pdf_x}.
Perturbing the evolution with random force kicks\footnote{
    Such that each well is tilted with a random force that we take independent and identically distributed (iid) according to a normal distribution with zero mean.
}
changes nothing to our observations, as included in \cref{fig:pdf_x} (see caption for markers).
We, furthermore, show that the probability density of the kinetic energy with which the particle enters the final well, $P(\mathcal{E})$, is constant as small argument in \cref{fig:E}.

\section{Concluding remarks}

Our central result is that $P(x) \sim x^2 P_w(x)$ in our toy model.
For a disorder $P_w(w) \sim w^k$ we thus find $P(x) \sim x^{k + 2}$.
We expect this result to qualitatively apply to generic depinning systems in the presence of inertia.
In particular, they are qualitatively (but not quantitatively) consistent with our previous empirical observations $\theta \simeq 2.5$ for $k = 1$ \cite{deGeus2019} and $\theta \simeq 1.4$ for $k = 0.2$.
A plausible limitation of our approach is underlined by the following additional observation: in Ref.~\cite{deGeus2019}, it was found that for $x$ to be small, the stopping well was typically small (by definition), but also that the next well had to be small.
Such correlations can exist only if the degree of freedom had visited the next well, before coming back and stopping.
This scenario cannot occur in our simple description where the particle only moves forward, except when it oscillates in its final well.
This implies that elastic interactions have to be considered in order to obtain a quantitative agreement with our previous observations.

\section*{Disclaimer}

The authors declare no competing interests.
Data and code are freely available, see dataset that includes relevant links to the code~\cite{deGeus2024_armouring_data}.

\section*{Acknowledgements}

T.G.~acknowledges support from the Swiss National Science Foundation (SNSF) by the SNSF Ambizione Grant PZ00P2{\_}185843.

\bibliographystyle{unsrtnat}
\bibliography{library}

\appendix

\section{Analytical solution}
\label{sec:proof:solution}

\paragraph{Ansatz}

We look for the solution of a general linear equation of motion in one well
\begin{equation}
    m \ddot{r} + \eta \dot{r} + \kappa r - F = 0 .
    \label{eq:motion:general}
\end{equation}
where the position $r$ is expressed relative to the local minimum in potential energy.
The external force $F$ tilts the potential energy landscape and is used as a perturbation to check the robustness of our argument.
An ansatz to this differential equation is
\begin{equation}
    r(\tau) = \alpha e^{-\lambda_- \tau} + \beta e^{-\lambda_+ \tau} + \Delta r,
    \label{eq:ansatz:r}
\end{equation}
where $\Delta r = F / \kappa$, and $\tau$ is the time that the particle has spent since the entry in the current well at $r(\tau = 0) \equiv - w / 2$.
We denote the particle's velocity $v(\tau) \equiv \dot{r}(\tau)$, whereby we take $v(\tau = 0) \equiv v_0$.
\detail{
    \begin{equation}
        \dot{r} (\tau) = - \alpha \lambda_- e^{- \lambda_- \tau} -
        \beta \lambda_+ e^{- \lambda_+ \tau}
    \end{equation}
    such that
    \begin{equation}
        \ddot{r}(\tau) = \alpha \lambda_-^2 e^{- \lambda_- \tau} +
        \beta \lambda_+^2 e^{- \lambda_+ \tau} .
    \end{equation}
    Substituting in \cref{eq:motion:general} gives
    \begin{equation}
        \begin{split}
            (m \lambda_-^2 - \eta \lambda_- + \kappa) \alpha e^{- \lambda_- \tau} + \\
            (m \lambda_+^2 - \eta \lambda_+ + \kappa) \beta e^{- \lambda_+ \tau} + \\
            \kappa \Delta r - F = 0
        \end{split}
    \end{equation}
    The last line cancels out, such that
    \begin{equation}
        \begin{split}
            (m \lambda_-^2 - \eta \lambda_- + \kappa) \alpha e^{- \lambda_- \tau} + \\
            (m \lambda_+^2 - \eta \lambda_+ + \kappa) \beta e^{- \lambda_+ \tau} = 0
        \end{split}
    \end{equation}
    which is satisfied for any $(\alpha, \beta)$ if
    \begin{equation}
        (m \lambda_-^2 - \eta \lambda_- + \kappa) = (m \lambda_+^2 - \eta \lambda_+ + \kappa) = 0
    \end{equation}
}
Substituting this ansatz in \cref{eq:motion:general} leads to $\lambda_\pm = (\eta \pm \sqrt{\eta^2 - 4 m \kappa}) / (2 m)$.
The prefactors $\alpha$ and $\beta$ are set by the initial conditions, such that
\detail{
    \begin{align}
        - w / 2 & = \alpha + \beta + \Delta r
    \end{align}
    so that
    \begin{align}
        \alpha + \beta = - w / 2 - \Delta r \equiv r_0
    \end{align}
    which gives
    \begin{align}
        (r_0 - \beta) \lambda_- + \beta \lambda_+ = - v_0 \\
        r_0 \lambda_- + \beta (\lambda_+ - \lambda_-) = - v_0 \\
        \beta = - \frac{r_0 \lambda_- + v_0}{\lambda_+ - \lambda_-} \\
        \alpha = r_0 - \beta = \frac{r_0 \lambda_+ + v_0}{\lambda_+ - \lambda_-}
    \end{align}
}
\begin{equation}
    \alpha, \beta = \pm \frac{r_0 \lambda_{\pm} + v_0}{\lambda_+ - \lambda_-} ,
    \label{eq:alpha}
\end{equation}
with $r_0 \equiv - w / 2 - \Delta r$.

\paragraph{Underdamped}

We recognise that if $\lambda_\pm$ are real ($\eta^2 > 4 m \kappa$), the dynamics are overdamped and the velocity decays exponentially.
Conversely, the underdamped dynamics that we consider correspond to $\eta^2 < 4 m \kappa$~\footnote{
    Note that our solution warrants some caution for critical damping $\eta^2 = 4 m \kappa$.
}.

\paragraph{Oscillator}

In the underdamped case, $\lambda_\pm$ and $\alpha, \beta$ are complex conjugates.
This allows us to simplify the solution by expressing those coefficients as $\lambda_\pm = \lambda \pm i \omega$ and $\alpha, \beta = (L / 2) e^{\pm i \phi}$ as follows~\footnote{
    $\cos(z) = (e^{i z} + e^{- i z}) / 2$
}
\begin{equation}
    r(\tau) = L e^{- \lambda \tau} \cos(\omega \tau + \phi) + \Delta r .
    \label{eq:solution:r}
\end{equation}
We remark that the velocity $v(\tau)$ can be expressed as a phase shift with respect to $r(\tau)$~\footnote{
    $a \cos(z) + b \sin(z) = \mathrm{sgn}(a) \sqrt{a^2 + b^2} \cos\left(z - \arctan(b / a)\right)$
}
\detail{
    \begin{equation}
        v(\tau) = - L e^{- \lambda \tau} \big( \lambda \cos(\omega \tau + \phi) + \omega \sin(\omega \tau + \phi) \big)
    \end{equation}
}
\begin{equation}
    v(\tau) = - A e^{- \lambda \tau} \cos\left(\omega \tau + \phi - \arctan\left(\frac{\omega}{\lambda}\right) \right)
    \label{eq:solution:v}
\end{equation}
with $A = \lambda L \sqrt{1 + (\omega / \lambda)^2}$.
We summarise the amplitudes, frequency, and phase.
From $\lambda_\pm$ we find
\begin{equation}
    \lambda = \frac{\eta}{2 m}, \quad \omega^2 = \frac{\kappa}{m} - \left(\frac{\eta}{2m}\right)^2.
    \label{eq:lambda_and_omega}
\end{equation}
\detail{
    \begin{equation}
        (\omega / \lambda)^2 = \frac{4 \kappa m - \eta^2}{\eta^2}
    \end{equation}
}
Furthermore, \cref{eq:alpha} gives
\detail{
    \begin{equation}
        \alpha, \beta = \pm \frac{(\lambda \pm i \omega) r_0 + v_0}{2i \omega}
    \end{equation}
}
\begin{equation}
    \alpha, \beta = \tfrac{1}{2} \big[ r_0 \mp i (\lambda r_0 + v_0) / \omega \big],
\end{equation}
such that
\begin{align}
    L^2
    & = 4 \big[\Re(\alpha)^2 + \Im(\alpha)^2 \big] \\
    & = \big[ (\omega r_0)^2 + (\lambda r_0 + v_0)^2 \big] / \omega^2,
\end{align}
and
\begin{align}
    \phi
    & = \chi \pi + \arctan\left(\Im(\alpha) / \Re(\alpha) \right) \\
    & = \chi \pi - \arctan\left(\lambda / \omega + v_0 / (\omega r_0) \right),
\end{align}
where $\chi$ depends on $\alpha$~\footnote{
    $\Re(\alpha) \geq 0 \rightarrow \chi = 0$.
    $(\Re(\alpha) < 0, \Im(\alpha) \geq 0) \rightarrow \chi = 1$.
    $(\Re(\alpha) < 0, \Im(\alpha) < 0) \rightarrow \chi = -1$.
}.

\section{Exiting well}
\label{sec:proof:Ec}

We will show that the minimum kinetic energy with which the particle needs to enter a well to be able to exit $\mathcal{E}_c \propto w^2$, whereby we consider $F = 0$.
The particle exits the well if ${v_0 > v_c}$.
$v_c$ thus corresponds to the case that $r(\tau_e) = w / 2 = - r_0$ for which $v(\tau_e) = v_e = 0$.
Let us make the ansatz that $v_0 = v_c = w / \tau_c$ and look for the solution of $\tau_c$.

We note that on the interval $\tau \in [0, \tau_e)$ the position is strictly monotonically increasing.
The solution of $v(\tau_n) = 0$ corresponds to
\begin{equation}
    \omega \tau_n + \phi - \arctan(\omega / \lambda) = (n + 1/2) \pi, \quad n \in \mathbb{Z}.
    \label{eq:taue}
\end{equation}
\detail{
    \begin{equation}
        \omega \tau_n + \phi = (n + 1/2) \pi + \arctan(\omega / \lambda)
    \end{equation}
}
The for us relevant solution is $\tau_e = \min (\tau_n > 0)$~\footnote{
    i.e.\ $n = \pm 1$ depending on $(r_0, v_0)$
}.
\detail{
    \begin{equation}
        \alpha, \beta = (w / 4) \left( - 1 \mp i \left( \frac{2 / \tau_c - \lambda}{\omega} \right) \right),
    \end{equation}
    such that
    \begin{equation}
        \Re(\alpha) < 0
    \end{equation}
    such that $\chi = \text{sign} \left( 2 / \tau_c - \lambda \right)$ and
    \begin{equation}
        \phi = \chi \pi + \arctan\left( \big[ 2 / \tau_c - \lambda \big] / \omega \right),
    \end{equation}
}
$r(\tau_e) = w / 2$ corresponds to
\detail{
    \begin{equation}
        r(\tau_e) = L e^{- \lambda \tau_e} \cos((n + 1/2) \pi + \arctan(\omega / \lambda)) = w / 2
    \end{equation}
}
\begin{equation}
    L e^{- \lambda \tau_e} = w / (2 c_0) ,
\end{equation}
with
\begin{equation}
    c_0 = \cos((n + 1/2) \pi + \arctan(\omega / \lambda)) .
\end{equation}
Using the definition of $\tau_e$ in \cref{eq:taue} leads to
\detail{
    \begin{equation}
        L e^{- \lambda (n + 1/2) \pi / \omega + \lambda \phi / \omega - (\lambda / \omega) \arctan(\omega / \lambda)} = w / (2 c_0)
    \end{equation}
}
\begin{equation}
    L e^{\lambda \phi / \omega} = w / (2 c_0 c_1) ,
\end{equation}
with
\begin{equation}
    c_1 = e^{- \lambda (n + 1/2) \pi / \omega - (\lambda / \omega) \arctan(\omega / \lambda)}.
\end{equation}
Furthermore,
\begin{equation}
    L^2 = \tfrac{1}{4} \big[ \omega^2 + (2 / \tau_c - \lambda)^2 \big] (w / \omega)^2,
\end{equation}
\begin{equation}
    \phi = \text{sign} \left( 2 / \tau_c - \lambda \right) \pi + \arctan\left( \big[ 2 / \tau_c - \lambda \big] / \omega \right),
\end{equation}
such that
\begin{equation}
    L' e^{\lambda \phi / \omega} = 1 / (2 c_0 c_1) ,
    \label{eq:tauc}
\end{equation}
with $L' = L / w$ (such that $L'$ is $w$ independent).
\detail{
    \begin{equation}
        L' = (1 / \omega) \sqrt{ (\omega / 2)^2 + (1 / \tau_c - \lambda / 2)^2 },
    \end{equation}
}
\cref{eq:tauc} is $w$ independent and can be solved for $\tau_c = \tau_c(\lambda, \omega, n)$, proving that $v_c = w / \tau_c$.
This result thus corresponds to
\begin{equation}
    \mathcal{E}_c = \frac{1}{2} m v_c^2 = \frac{m}{2 \tau_c^2} w^2.
\end{equation}
as we used to go from \cref{eq:pstop} to obtain \cref{eq:rhos} using \cref{eq:pec}.

\section{Entry kinetic energy}
\label{sec:proof:Kw}

With $\mathcal{K}$ the kinetic energy at exiting the well, we show that $\partial_w \mathcal{K} \big|_{w = w^*} > 0$.
We again consider $F = 0$.
In particular, we show that
\begin{equation}
    \partial_w \mathcal{K} = m \, v_e \, \partial_w v_e > 0 .
\end{equation}
The derivative of the velocity as a function of the well size corresponds to
\begin{equation}
    \partial_r v = \partial_\tau v / \partial_\tau r = a(\tau) / v(\tau) ,
\end{equation}
where the acceleration $a(\tau) \equiv \ddot{r}(\tau) = \alpha \lambda_-^2 e^{- \lambda_- \tau} + \beta \lambda_+^2 e^{- \lambda_+ \tau}$.
By evaluating this expression with initial conditions $r(\tau = 0) = - w^* / 2$ and $v(\tau = 0) = 0$, we find
\begin{equation}
    \partial_w \mathcal{K} \big|_{w = w^*} = m \, (\alpha \lambda_-^2 + \beta \lambda_+^2) = - \frac{m}{2} (\lambda^2 + \omega^2) \, w^* .
\end{equation}
From the definitions of $\lambda$ and $\omega$ in \cref{eq:lambda_and_omega}, we thus find
\begin{equation}
    \partial_w \mathcal{K} \big|_{w = w^*} = - \kappa w^* / 2 ,
\end{equation}
as we argued above to show that $P(\mathcal{E} = 0) > 0$ using \cref{eq:rhoe0}.

\end{document}